# Epitaxial $MgB_2$ thin films on $ZrB_2$ buffer layers: structural characterization by synchrotron radiation

V.Ferrando[a], C.Tarantini[a], E.Bellingeri[a], P.Manfrinetti[b], I.Pallecchi[a], D.Marré[a], O.Plantevin[c], M.Putti[a], R.Felici[d] and C.Ferdeghini[a]

[a] *INFM-LAMIA, Dipartimento di Fisica, via Dodecaneso 33, 16146 Genova, Italy*

[b] *INFM-LAMIA, DICCI, via Dodecaneso 31, 16146 Genova, Italy*

[c] *ID01-ESRF 6, rue J. Horowitz, BP220 38043 Grenoble, France*

[d] *INFM-OGG – ESRF 6, rue J. Horowitz, BP220 38043 Grenoble, France*

## Abstract

Structural and superconducting properties of magnesium diboride thin films grown by pulsed laser deposition on zirconium diboride buffer layers were studied. We demonstrate that the $ZrB_2$ layer is compatible with the $MgB_2$ two step deposition process. Synchrotron radiation measurements, in particular anomalous diffraction measurements, allowed to separate $MgB_2$ peaks from $ZrB_2$ ones and revealed that both layers have a single in plane orientation with a sharp interface between them. Moreover, the buffer layer avoids oxygen contamination from the sapphire substrate. The critical temperature of this film is near 37.6 K and the upper critical field measured at Grenoble High Magnetic Field Laboratory up to 20.3 T is comparable with the highest ones reported in literature.

## Introduction

Since the discovery of superconductivity in magnesium diboride [1], many efforts have been made to produce high quality thin films with structural and superconducting properties similar to those of single crystals. This is desirable both for fundamental studies and for applications in electronic devices. Due to the volatility of magnesium, the development of particular deposition techniques became necessary (for a review, see [2]); the best results have been obtained by a direct synthesis by Hybrid Physical Chemical Vapor Deposition [3], yet the two step technique first proposed in [4] is widely used. The more diffused substrates are in general oxide single crystals, in particular sapphire, as well as silicon carbide, which is on the other hand very expensive. During the annealing at high temperature (800-900°C) in the two step method, these oxide substrates begin to release oxygen. It has been recently observed [5] that during the high temperature heat treatment an epitaxial interlayer of MgO is formed between sapphire and $MgB_2$, thus affecting its superconducting properties. The use of another diboride of the family M-$B_2$, isostructural with $MgB_2$, can be very helpful to prevent oxygen contamination, as well as to induce in plane ordered growth. These diborides can be grown epitaxially at high temperature with standard PLD techniques, with good surface morphology, thus being promising candidates to be employed as substrates for magnesium diboride. Moreover, it has been shown [6,7] that in thin films the critical temperature can be higher than in bulk and single crystal samples; the tensile strain of the $MgB_2$ lattice due to the substrate causes a softening of $E_{2g}$ phonon and thus an increase in $T_c$ up to 41.8 K [8]. Up to now, this strain of *a* axis in $MgB_2$ has been obtained only using substrates with a small lattice mismatch, but in principle the enhancement of $T_c$ could be even higher if a substrate with appropriate crystallographic parameters is used. Actually, in the family of diborides, there are compounds with *a* parameter larger than $MgB_2$ ($ZrB_2$, $a$=3.169Å and $ScB_2$, $a$=3.147Å) [9], that can be used as buffer layers to observe a raise in $T_c$. In this paper, we present our first results concerning the growth of magnesium diboride thin films on zirconium diboride buffer layers. A complete structural characterization of the two layers, investigated by synchrotron radiation measurements, is widely described, together with the superconducting properties of the samples.

**Sample preparation and characterization**

Both the film layers of this work were deposited by Pulsed Laser Ablation technique in an ultra high vacuum chamber (minimum pressure of $10^{-10}$ mbar) equipped with a Reflection High Energy Electron Diffraction system for the direct study of the growth. Details about the experimental setup have been reported elsewhere [10]. Zirconium diboride layers were grown starting from a commercial dense and stoichiometric pellet (MTI components) on c-cut sapphire substrates at temperatures between 850 and 950 °C. The laser frequency and the beam fluency were kept constant at 3 Hz and 3 J/cm$^2$ respectively; with these parameters, the growth rate is around 0.03 Å/laser pulse. The RHEED analysis performed during deposition revealed a three dimensional ordered growth, with a single in plane orientation. The spacing between the spots changes when the sample is rotated by 30 degrees with respect to the electron beam and the ratio between the distances is 1.7, which is the expected value for a hexagonal system. The $MgB_2$ films were grown by the two step technique previously described in [11]. The amorphous stoichiometric precursor layer was deposited at room temperature on the $ZrB_2$ film in the same vacuum conditions without opening the chamber. Afterward, they were annealed ex situ at 875°C in magnesium vapor, so to obtain the superconducting phase. The structural characterization of this kind of samples has been carried out using synchrotron radiation at ID01 beam line at ESRF, Grenoble. This was necessary because of the very similar lattice parameters and crystal structures of magnesium and zirconium diboride which yields superimposed diffraction peaks and also because of their low scattering power. The measurements were performed with a six-circle diffractometer which allowed to explore the reciprocal space in different geometries. The reflectivity and diffraction measurements were performed at beam energy of 11 keV. At this energy, the signal coming from $ZrB_2$ phase was predominant. To resolve $MgB_2$ peaks Anomalous X-ray Diffraction (AXD) was used. This particular technique is based on the energy dependence of the x-ray form factor of an element close to an absorption edge. Setting the beam energy near this absorption edge, it is possible to obtain a contrast in the scattering of that element compared to other elements in the sample. Therefore, to distinguish $MgB_2$ layer from the buffer layer, we measured the diffraction from the sample also at 17.93 KeV, near the absorption edge of Zr, so that the intensity of $ZrB_2$ peaks is suppressed and it is possible to distinguish the two phases. This energy value has been estimated measuring the intensity of a diffraction peak as a function of the beam energy: near the absorption edge, the intensity has a minimum, and thus energy corresponding to this minimum has been chosen. The same behavior has been found also using the fluorescence of a zirconium foil. Atomic Force Microscopy measurements in contact mode were carried out in order to study the morphology of the samples. From the electrical point of view, critical temperature and magnetoresistance were measured at GHMFL in Grenoble by standard four probe technique in an applied magnetic field up to 20.3 T both parallel and perpendicular to the *ab* planes.

**Results and discussion**

In the left panel of figure 1, classical Bragg-Brentano geometry scattering starting from an angle smaller than critical angle for total reflection is shown: this geometry allows measuring both the reflectivity and the diffraction pattern from the sample. In the low angle region, magnified in figure 2, very intense oscillations in the scattered intensity have been detected, indicating that the interfaces between the layers are sharp. Looking carefully to this measurement, two different families of oscillations are present (see inset of figure 2 where the small angle part of the scan is enlarged). Oscillations of larger periodicity, well evident up to $q=0.6$ Å$^{-1}$, can be ascribed to $ZrB_2$ buffer layer while oscillations of smaller periodicity, disappearing at lower q values, are attributed to the thicker $MgB_2$ film. The thickness and the surface roughness of the two films, evaluated from the fitting of the curve, which is plotted as dashed line in figure 1, are: $d=84$ Å and $\sigma=4$ Å for zirconium diboride and $d=1120$ Å and $\sigma=40$ Å for magnesium diboride. This roughness value is compatible with those of 55 Å obtained by Atomic Force Microscopy measurements in contact mode on the $MgB_2$ film on areas of 1 μm x 1 μm, among the droplets. An image on a larger area of the sample is reported in figure 3, panel A. It revealed a quite good quality of the sample in the regions among the particulates coming from the target during the ablation process. Also previous studies of the morphology of similar $ZrB_2$ films by AFM (panel B of fig.3) showed a RMS roughness of around 6 Å, compatible with our estimation by reflectivity measurements.

In the left panel of figure 1, the entire diffraction pattern of the multilayer is reported. Only two intense peaks are present, ascribed to the 00*l* reflections of the more crystalline and more diffracting (due to the larger atomic number of Zr) $ZrB_2$ film, the $MgB_2$ peaks being probably hidden. As already mentioned indeed, the angular positions of $MgB_2$ and $ZrB_2$ diffraction peaks are very close to each other due to the very similar structure and crystallographic parameters. On these peaks, also finite size effects are visible, with clear oscillations on both sides of 001 and 002 reflections. By fitting the finite size fringes on 001 peak (figure 1 left panel) after a convolution of the data with an instrumental resolution around 2%, no displacement in vertical direction has been observed. Vertical displacement, in fact, does not influence the minimum of the oscillations but only the maximum. Moreover, the thickness calculated from the fit is of the order of 85 Å, which is fully compatible with that evaluated from reflectivity measurements. To take into account the roughness of the $ZrB_2$ film, the fit has been obtained considering occupancy of 0.5 and 0.9 for the uppermost surface cell and for the remaining cells respectively: these values are in good agreement with those obtained for the roughness of $ZrB_2$ by reflectivity measurements. Therefore, we can conclude that no amorphous or reacted layer is present neither between the substrate and the $ZrB_2$ film nor between the buffer layer and the $MgB_2$ film: this is an interesting result because it gives a confirmation that diborides buffer layers are compatible with the growth of magnesium diboride, also by a two step method. From the angular position of the 00*l* reflections, we calculated the *c* axis for the $ZrB_2$ film: a value of 3.573±0.003 Å has been obtained, considerably higher than the bulk one (3.530 Å [9]). The strong orientation of the multilayer along the *c* axis has been confirmed by the rocking curve measurements around 00*l* reflections reported in figure 4. Observing these curves, it can be noted that there is the superimposition of two peaks: one very narrow, with FWHM of 0.15° and the other broad and large, much more evident in the 002 reflection, which shows a FWHM of around 1.6°. This suggests that the two diboride layers have different average misalignment along the *c* axis; the well crystallized $ZrB_2$ gives the narrow peak while the large part of the curve can be ascribed to $MgB_2$, which seems not perfectly oriented along the *c* axis, as already observed also on other substrates. The contribution of $MgB_2$ layer is more obvious in 002 peak because in the calculated intensity 002 reflection is three times more intense than 001, differently from $ZrB_2$, where the opposite situation occurs. The split of the narrow peak both in left and right panel of figure 4 is quite puzzling. Our hypothesis is that it could be due to a well crystallized part of $MgB_2$ phase which is present in the film, as already observed in [12]. This suggestion is also strengthened by the fact that the relative intensity of the two maxima is swapped in 001 and 002 rocking curves, consistently with the calculated intensity for $ZrB_2$ and $MgB_2$.

In order to check the in plane orientation of the film, *phi* scan measurements on 101 reflection of zirconium diboride have been carried out and they are shown in figure 5. Only peaks spaced of 60° have been observed, that could be attributed both to $ZrB_2$ and $MgB_2$. Due to the absence of any other peak, the in plane orientation of magnesium diboride should be the same of $ZrB_2$. The orientation of the two layers with respect to the substrate is rotated of 30°, as clarified also by RHEED analysis of the $ZrB_2$ film during the growth; this alignment of the diboride film with respect to the $Al_2O_3$ results in fact in a lower mismatch between the two materials [13]. Actually, the peaks in the *phi* scan are very broad, up to 8 degrees FWHM, indicating a marked in plane mosaicity in the sample. In the inset of figure 5, reciprocal space sections (L scans) around one peak of the phi scan (position indicated A in the figure) and in between two peaks (B) are shown in units of the crystallographic parameters of the sapphire substrate. In the following, all the scans in H, K or L will be reported in these units. L scanning around the position of the peak can allow to resolve the $MgB_2$ contribution, while by the L scan in B it is possible to check if there is a fraction of the sample not perfectly in plane oriented. The curve around A shows a small shoulder on the right part of the peak, that can be ascribed to $MgB_2$, but it could not be more evidenced by these measurements. On the other hand, the scan in B presents a weak peak, with an intensity of more than one order of magnitude lower than in A; this indicates that there is a very small part of the sample, $MgB_2$ or $ZrB_2$, with a random in plane orientation with respect to the substrate, but it is anyway negligible if compared with the oriented one. To distinguish the contributions of the two diboride phases, Anomalous X-ray Diffraction measurements (AXD) have been carried out. A K scan around 101 peak at two different energies is reported in figure 6; varying the incident beam energy near the zirconium absorption edge, the $ZrB_2$ peak is weakened, thus allowing evidencing the $MgB_2$ phase. From the FWHM obtained by fitting the $ZrB_2$ peak and the $MgB_2$ shoulder with two Gaussian curves, we calculated the in plane nanostructuration of the two layers: values of 230 and 240 Å for $MgB_2$ and $ZrB_2$ have been obtained respectively. At energies far from the zirconium absorption edge, no dependence on the energy is visible on the relative intensities of the peaks, and so it is possible to compare measurements done at very different energies, such as 11 KeV and 17.93 KeV. From the FWHM of L sections performed in the same conditions of K scan in fig. 6, the grain dimension along the *c* axis of the two phases resulted to be around 65 Å, not too different from that obtained by finite size fringes on 00l reflections for $ZrB_2$. The small difference can be ascribed to a gliding of the planes, which cannot be evidenced by diffraction from 00*l* peaks.

Reciprocal space mapping is a very powerful tool to have a two-dimensional image of the reciprocal space; moreover, making sections of these maps in H, K or L directions it is possible to clarify the structure of the peak and to calculate the crystallographic parameters. Figure 7 shows a comparison between reciprocal space maps in HK space around 101 reflection in standard (left panel, E=11 KeV) and AXD mode (right panel, E=17.93 KeV). Looking at the graph in the left panel, a single intense peak, in the HK coordinates expected for $ZrB_2$, is well evident. Anyway, an asymmetry in its shape can be noted, suggesting that it is the convolution of two peaks. Sections of the map around the peak in H and K directions at the different L values of zirconium and magnesium diboride, similar to those of fig.6, confirmed our hypothesis, revealing that the peak has a complex structure, with a broadening in the right side, probably related to $MgB_2$. The asymmetry is much more evident in the map shown in right panel, where the incident energy is near the Zr absorption edge: in this case, the intensity of $ZrB_2$ peak is depressed, and the $MgB_2$ one clearly appears. It should be noted that the magnesium diboride peak has a halo with not zero intensity on its sides that can be ascribed to a randomly oriented part of the superconducting phase, as already observed in the inset B of figure 5. The cuts of the peak in the reciprocal space have been performed also on this measurement: from H and K scans along the peaks, the *a* axis of the layers can be estimated while the *c* axis can be calculated from L scans. In table I average crystallographic parameters values for $ZrB_2$ and $MgB_2$ calculated from various measurements are reported, together with the bulk values, as a reference. The *a* axis of $ZrB_2$ is significantly reduced with respect to the bulk value, and consequently *c* axis is enlarged, in order to keep the cell volume constant. In the case of $MgB_2$ film, the in plane lattice parameter is, within the error bar, similar to the bulk one, while the *c* axis is considerably extended. This suggests that the strain produced by the buffer layer is small in this case. By this growth method in fact, the crystallization of the $MgB_2$ does not occur during the growth, when it would be most sensitive to the underlying layer. The buffer layer certainly helps epitaxy of the superconducting film, but probably it is not able to induce a lattice strain in the ex situ crystallized $MgB_2$ film. As a consequence of the lack of a strong in plane lattice strain, no enhancement of critical temperature above the bulk value has been observed in this sample, even if the measured value of 37.6 K (90% of normal state resistivity, see inset of figure 8) is quite high if compared with standard values obtained by our two step technique. Resistivity in the normal state (40K) of the film resulted to be 12 μΩcm, very low for a sample prepared by a two steps technique. In the calculation of ρ, a typical resistivity value for the $ZrB_2$ film (1 mΩcm) in parallel to $MgB_2$ has been considered.

In a recent work, V.Braccini et al. [14] observed a correlation between the rising of the *c* axis and the upper critical field enhancement in a set of thin films prepared by very different techniques.

Furthermore, it has been noted that samples with very low resistivity can show very high critical fields, differently from what expected in BCS [16] and this is a peculiarity of two bands systems. The good structural properties, the extended *c* axis and the very low resistivity make the sample of this paper a good candidate for studying $H_{c2}$. Magnetoresistivity measurements parallel and perpendicular to the *ab* planes have been carried out up to 20.3 T at GHMFL in Grenoble: $H_{c2}(T)$ curves, calculated as the 90% of normal state resistivity, are shown in figure 8. Their linear shape at low temperatures is well evident, especially in perpendicular direction, where no saturation down 2K has been observed. Also applying the magnetic field parallel to the planes, the curve is quite linear, at least up to 20.3 T, the maximum field applied, thus allowing a linear extrapolation of $H_{c2}$ values. The obtained values of 17 and 48 T in perpendicular and parallel orientations respectively are comparable to the highest recently reported in literature for thin films [14, 15]. The anisotropy, calculated as the ratio between upper critical field parallel and perpendicular to the basal planes and reported in figure 8 as an inset, resulted to be around 3 at 20 K, decreasing with increasing temperature, thus suggesting that the disorder can be especially effective in sigma band [17]. These measurements confirmed that there is a correlation between the high $H_{c2}$ values and the expanded *c* axis, even in case of low resistivity samples. The physical meaning of this phenomenological observation is still lacking. Gurevich et al. [18] hypothesized that the increase of the *c* parameter can be ascribed to the buckling of Mg planes, as evidenced by TEM measurements in alloyed thin films [18]. On the other hand, Pogrebnyakov et al. [19] found a strong enhancement of the *c* axis in carbon doped thin films, up to 30%at of Carbon; in this case, TEM analyses indicate the presence of an amorphous C-rich phase between $Mg(B_{1-x}C_x)_2$ columnar nanograins. The sample presented here was prepared starting from stoichiometric precursor without carbon addition, but anyway Braccini et al. [14] showed that carbon is often present in $MgB_2$ thin films, especially when crystallized during an ex situ annealing.

**Conclusions**

In this paper, we showed that epitaxial zirconium diboride buffer layers are good substrates for magnesium diboride growth. Anomalous X-ray Diffraction measurements performed using synchrotron radiation allowed to distinguish the two diboride phases, indicating that both have a single in plane orientation rotated of 30° with respect to the sapphire substrate. Moreover, $ZrB_2$ resulted to be an optimal barrier for the oxygen diffusion from the $Al_2O_3$ at high temperatures: in fact, differently from what observed in samples without buffer layer, no interlayer of MgO has been found. Although the tensile strain of the *a* axis produced by the $ZrB_2$ on $MgB_2$ film was not appreciable in our case, we think that using a cleaner deposition technique for the superconducting layer this effect should be observed. The high critical field measured on this sample, in which the *c* axis is larger than the bulk value, confirmed a possible correlation between these two properties, also in films with low electrical resistivity like the one discussed in this work.

**Acknowledgements**

This work was supported by the European Community through “Access to Research Infrastructure action of the Improving Human Potential Programme“. We acknowledge I. Sheikin for the assistance during our magnet time at Grenoble High Magnetic Field Laboratory.

.

**References**

[1] J. Nagamatsu, N. Nakawaga, T. Muranaka, Y. Zenitani and J. Akimitsu, Nature 410 (2001) 63

[2] M. Naito and K. Ueda, Superc.Sci. and Technol. 17, (2001) R1

[3] X. Zeng, AV. Pogrebnyakov, A. Kothcharov, J.E. Jones, X.X. Xi, E.M. Lysczek, J.M. Redwing, S. Xu, Qi Li, J. Lettieri, D.G. Schlom, W. Tian, X. Pan and Z.K-Liu, Nature Materials 1, (2002) 35

[4] W.N. Kang, Hyeong-Jin Kim, Eun-Mi Choi, C.U. Jung, Sung-Ik Lee, Science 292 (2001) 1521

[5] V. Ferrando, C. Tarantini, E. Bellingeri, R. Felici, P. Manfrinetti and C. Ferdeghini, in preparation

[6] N. Hur, P.A. Sharma, S. Guha, Marta Z. Cieplak, D.J. Werder, Y. Horibe, C.H. Chen and S.-W. Cheong, Appl.Phys.Lett. 79 (2001) 4180

[7] X.H. Zeng, A.V. Pogrebnyakov, M.H. Zhu, J.E. Jones, X.X. Xi, S.Y. Yu, E. Wertz, Qi Li, J.M. Redwing, J. Lettieri, V. Vaithyanathan, D.G. Schlom, Zi-Kui Liu, O. Trithaveesak and J. Schubert, Appl.Phys.Lett. 82, (2003) 2097

[8] A.V. Pogrebnyakov, J.M. Redwing, S. Raghavan, V. Vaithyanathan, D.G. Schlom, S.Y. Xu, Qi Li, D.A. Tenne, A. Soukiassian, X.X. Xi, M.D. Johannes, D. Kasinathan, W.E. Pickett, J.S. Wu and J.C.H. Spence, to be published.

[9] P. Villars, L.D. Calvert, Pearson’s Handbook of Crystallographic Data for Intermetallic Phases, 2nd ed., ASM International, Materials Park, OH (1991)

[10] M.R. Cimberle, C. Ferdeghini, G. Grassano, D. Marré, M. Putti, A.S. Siri, F. Canepa, IEEE Trans. Appl. Supercond., 9 (1999) 1727

[11] V. Ferrando, S. Amoruso, E. Bellingeri, R. Bruzzese, P. Manfrinetti, D. Marrè, N. Spinelli, R. Velotta, X. Wang, C. Ferdeghini, Supercond.Sci. and Technol. 16(2003) 241

[12] C. Ferdeghini, V. Ferrando, G. Grassano, W. Ramadan, E. Bellingeri, V. Braccini, D. Marré, M. Putti, P. Manfrinetti, A. Palenzona, F. Borgatti, R. Felici, L. Aruta, Physica C, 378(2002) 56

[13] S.D. Bu, D.M. Kim, J.H. Choi, J. Giencke, S. Patnaik, L. Cooley, E.E. Hellstrom, D.C.Larbalestier, C.B. Eom, J. Lettieri, D.G. Schlom., W. Tian, X. Pan, Appl. Phys. Lett. 81 (2002) 1851

[14] V. Braccini, A. Gurevich, J.E. Giencke, M.C. Jewell, C.B. Eom, D.C. Larbalestier, A. Pogrebnyakov, Y. Cui, B.T. Liu, Y.F. Hu, J.M. Redwing, Qi Li, X.X. Xi, R.K. Singh, R. Gandikota, J. Kim, B. Wilkens, N. Newman, J. Rowell, B. Moeckly, V. Ferrando, C. Tarantini, D. Marré, M. Putti, C. Ferdeghini, R. Vaglio, E. Haanappel, submitted to Phys. Rev. Letters

[15] V. Ferrando, P. Manfrinetti, D. Marré, M. Putti, I. Sheikin, C. Tarantini, C. Ferdeghini Phys. Rev.B 68, (2003) 094517

[16] N.R. Werthamer, E. Helfand, and P.C. Hohenberg, Phys. Rev. 147, (1966) 288

[17] A. Gurevich, Phys.Rev.B 67, (2003)184515

[18] A. Gurevich, S. Patnaik, V. Braccini, K.H. Kim, C. Mielke, X. Song, L.D. Cooley, S.D. Bu, D.M. Kim, J.H. Choi, L.J. Belenky, J. Gienke, M.K. Lee, W. Tian, X.Q. Pan, A. Siri, E.E. Hellstrom, C.B. Eom and D.C. Larbalestier, Supercond. Sci. Technol. 17, 278 (2004).

[19] A.V. Pogrebnyakov, X.X. Xi, J.M. Redwing, V. Vaithyanathan, D.G. Schlom, A. Soukiassian, S.B. Mi, C.H. Jin. J.E. Gienke, C.B. Eom, J. Chen, Y.F. Hu, Y. Cui and Qi Li, submitted to Appl.Phys. Lett.

**Figure caption**

**Figure 1**: Left panel: Diffraction in classical Bragg Brentano geometry. Right panel: fit of finite size oscillations in 001 peak (line).

**Figure 2**: Reflectivity measurement of the film (circles) and fit (line). In the inset, magnification of the low angle region, in which fringes due to $MgB_2$ film are visible.

**Figure 3:** AFM images of the $MgB_2$ film (panel A) and of a $ZrB_2$ film similar to the buffer layer of this work (B). The RMS roughness is 55 and 6 Å respectively.

**Figure 4**: Rocking curves around 001 (left panel) and 002 reflections (right panel).

**Figure 5**: Phi scan around 101 reflection. In the insets, L scans in the region called A and B in the figure.

**Figure 6**: K scan at different energies (17.93 KeV, full symbols, and 17.89 KeV, open symbols) around a $ZrB_2$ peak.

**Figure 7**: Reciprocal space maps around 101 reflection at incident energy of 11 KeV (left panel) and in anomalous diffraction conditions with E=17.9 KeV (right panel).

**Figure 8**: Upper critical field parallel (open symbols) and perpendicular (full symbols) orientation up to 20.3 T. In the inset, resistance and anisotropy versus temperature curves.

**Table I**: Calculated crystallographic parameters of $ZrB_2$ and $MgB_2$ layers and for the bulk [9], as comparison.

| | *a axis Å* | *c axis Å* |
|---|---|---|
| $ZrB_2$ | 3.142±0.006 | 3.568±0.007 |
| Bulk $ZrB_2$ | 3.169 | 3.530 |
| $MgB_2$ | 3.090±0.006 | 3.545±0.007 |
| Bulk $MgB_2$ | 3.086 | 3.524 |

Table I

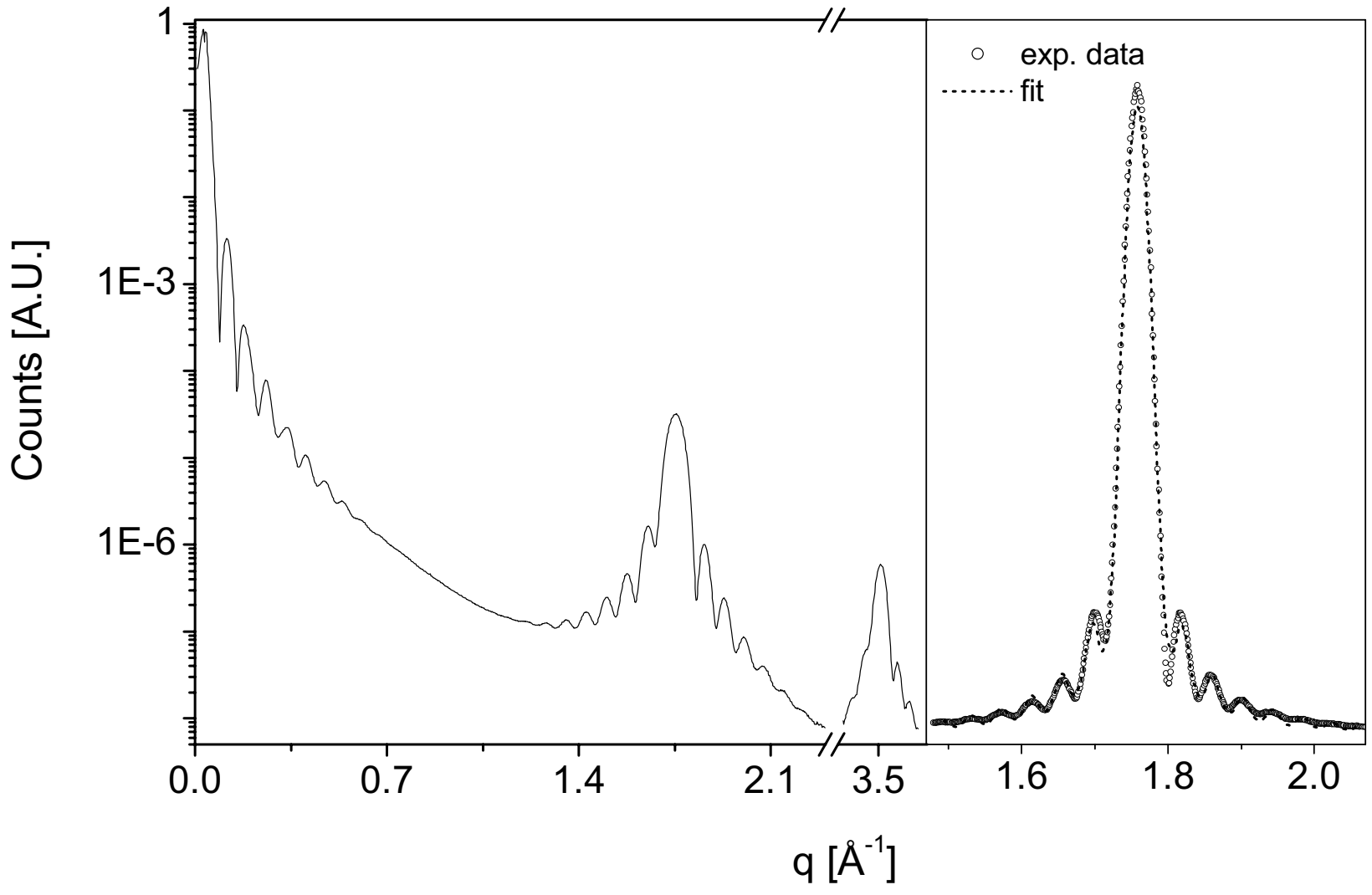

Fig.1

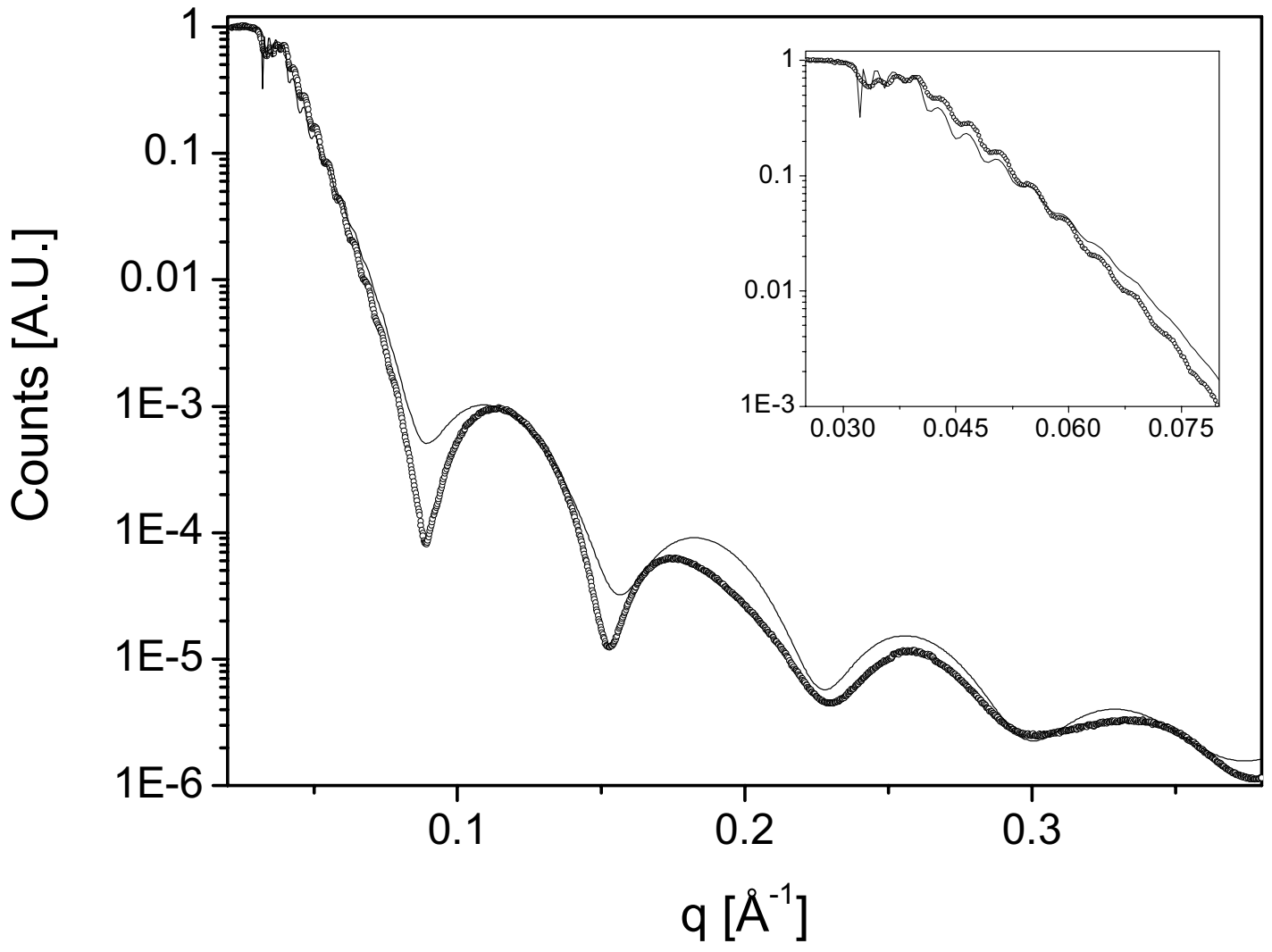

Fig.2

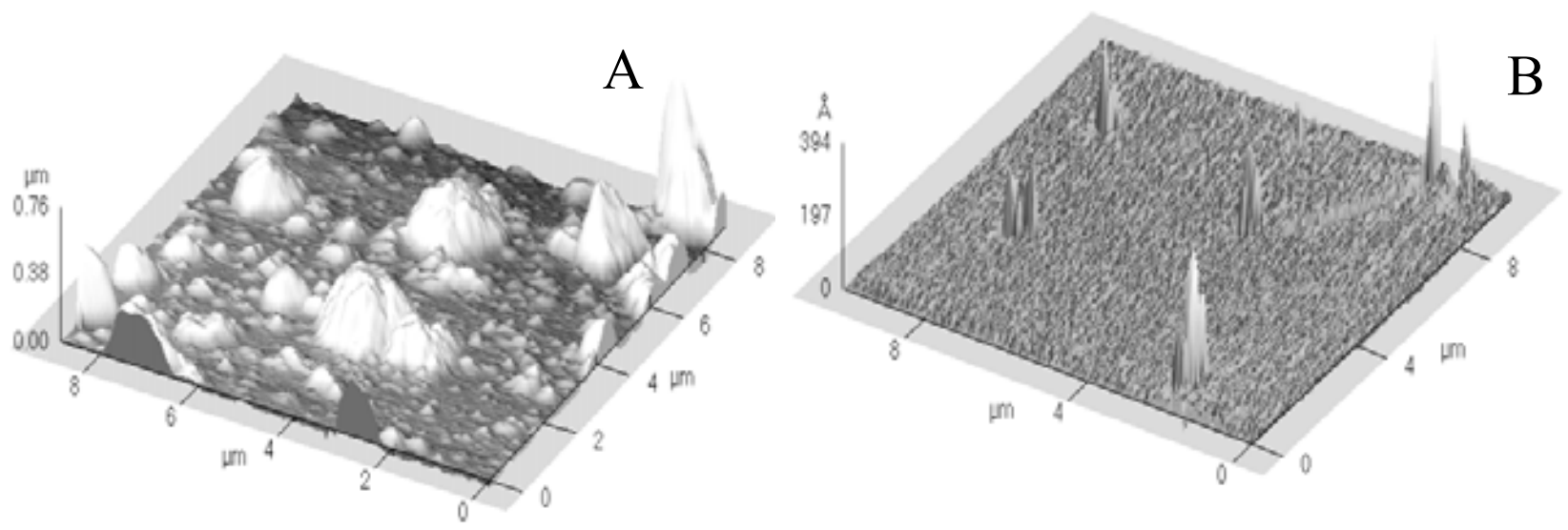

Fig.3

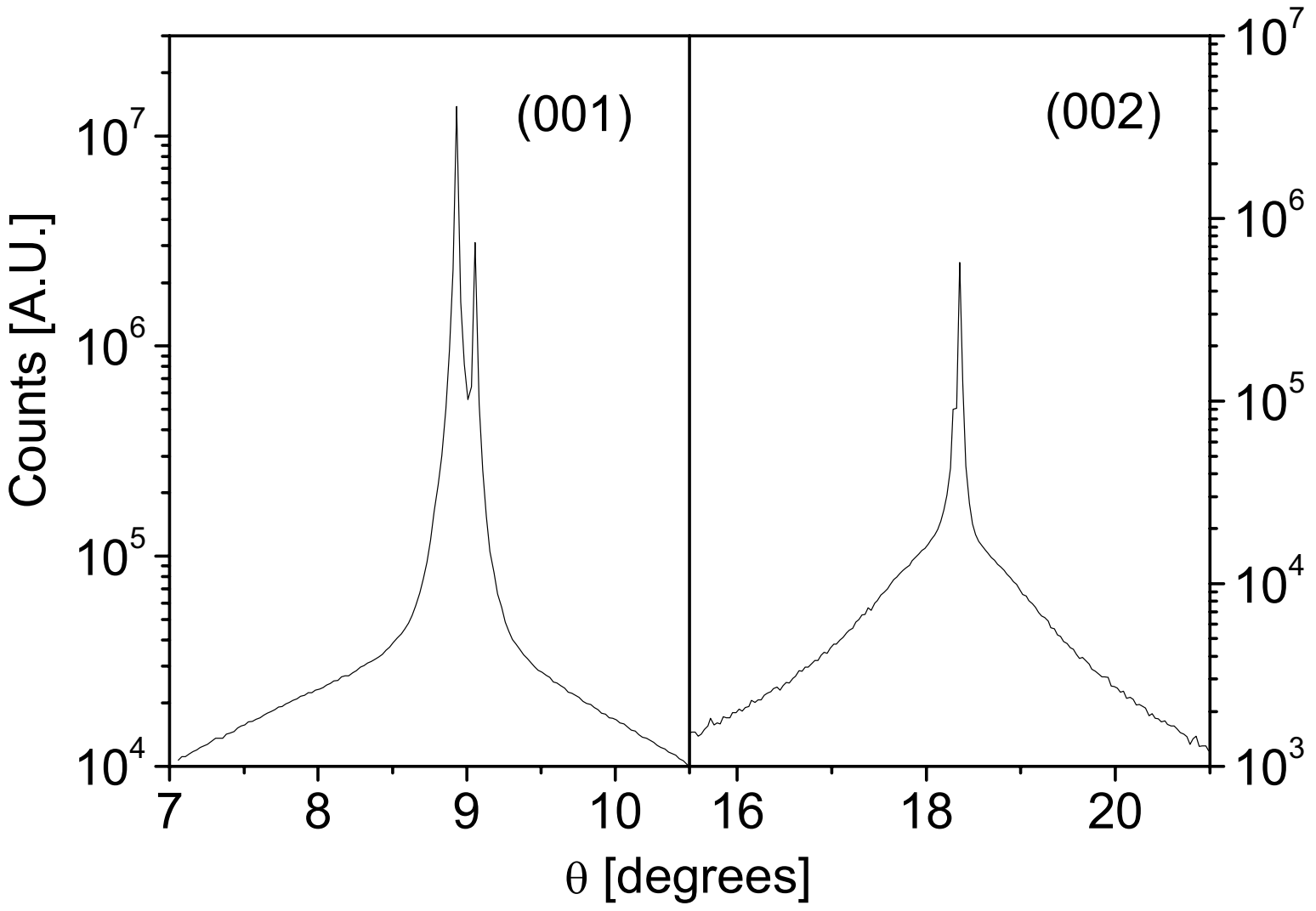

Fig.4

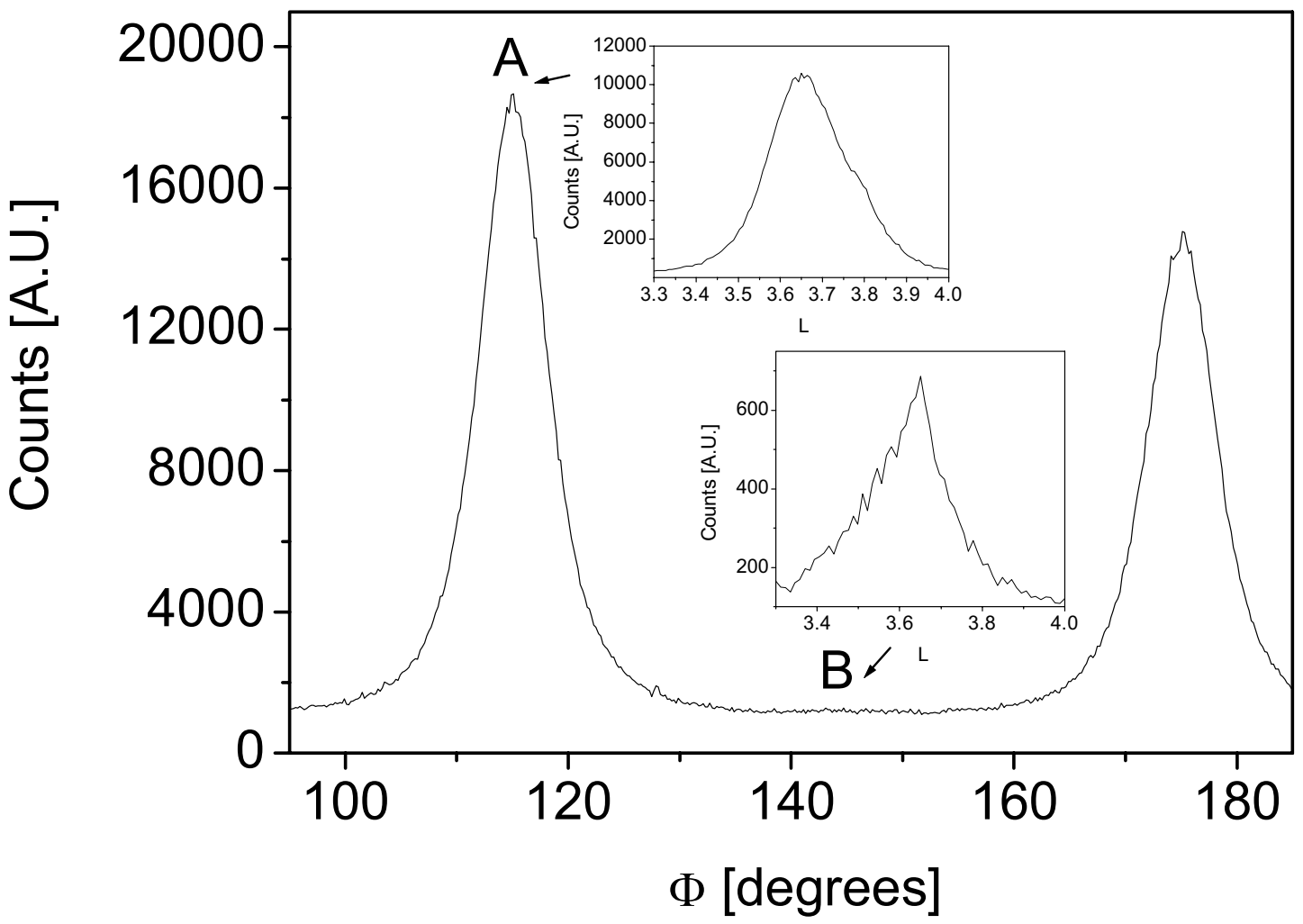

Fig.5

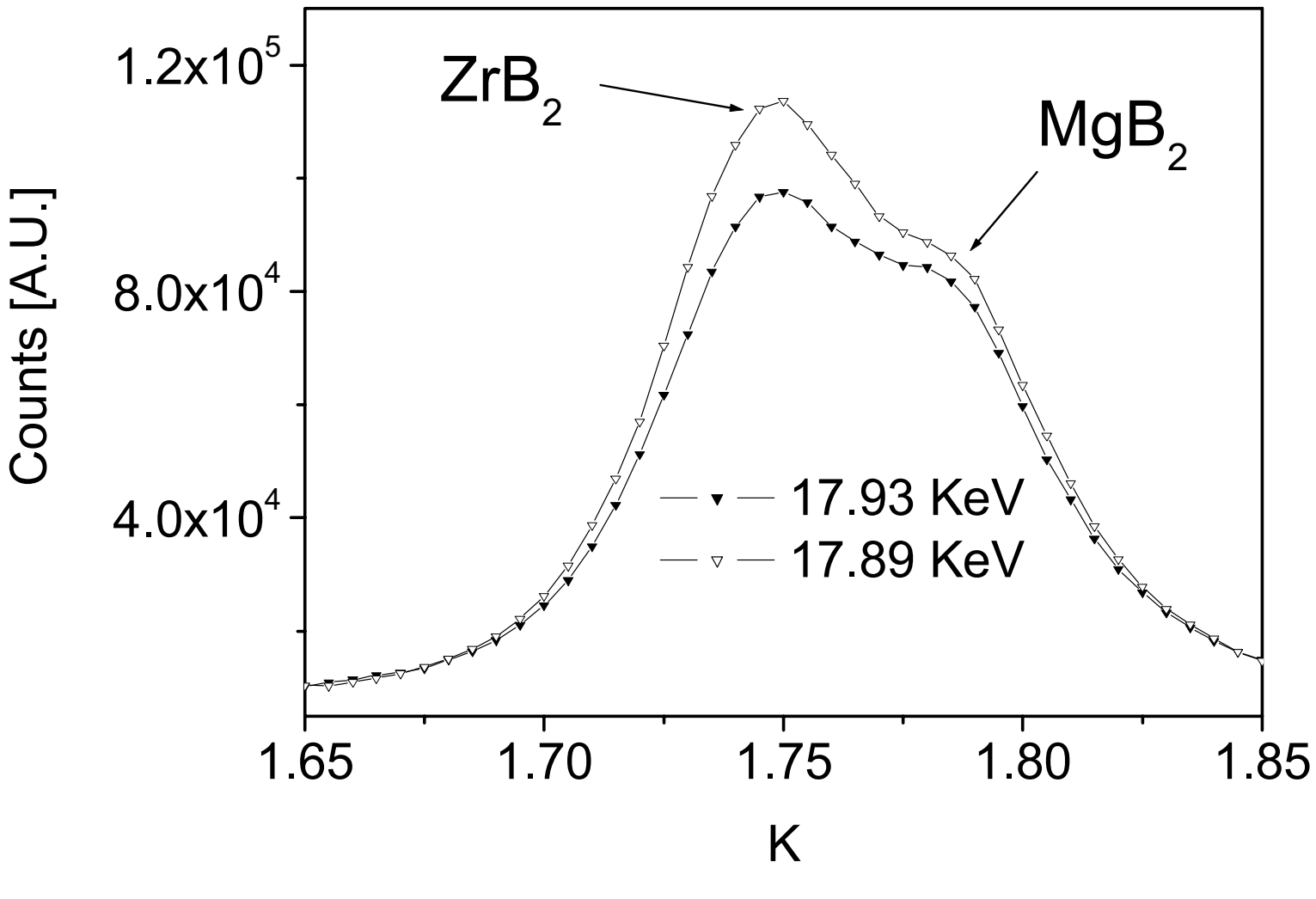

Fig.6

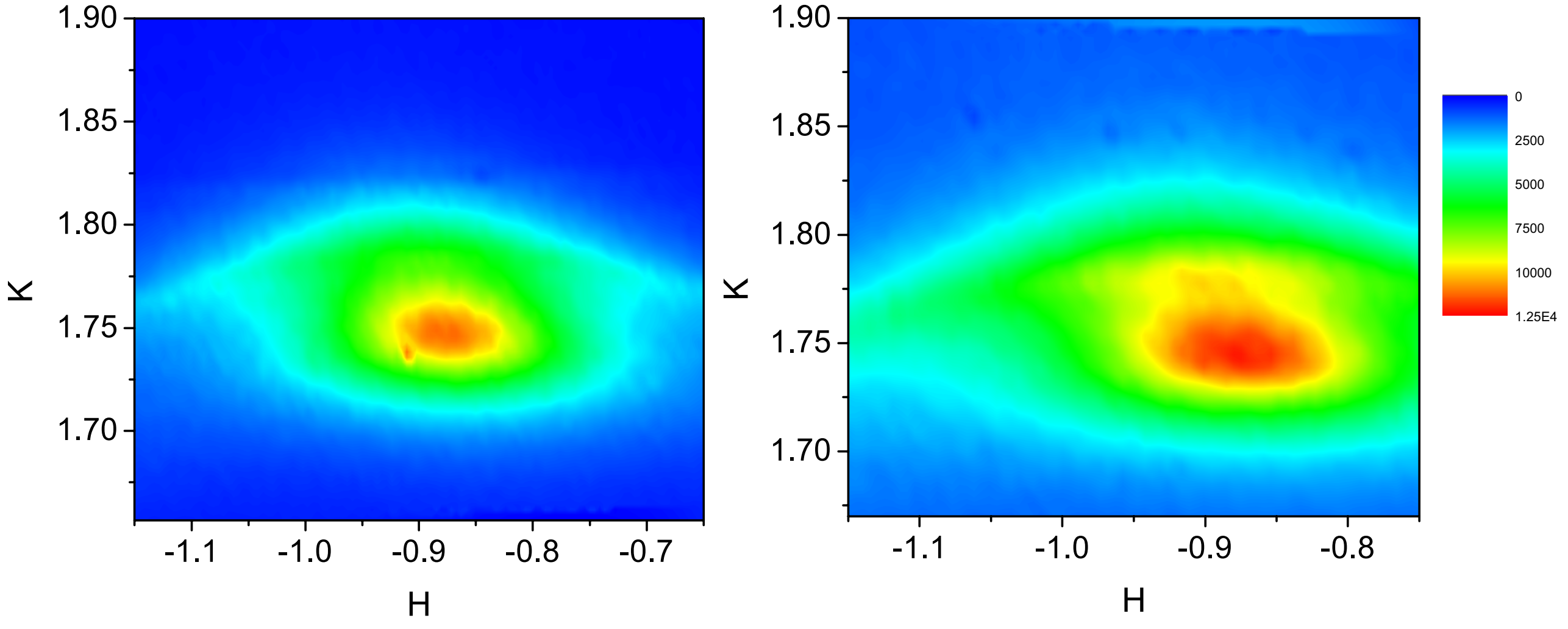

Fig.7

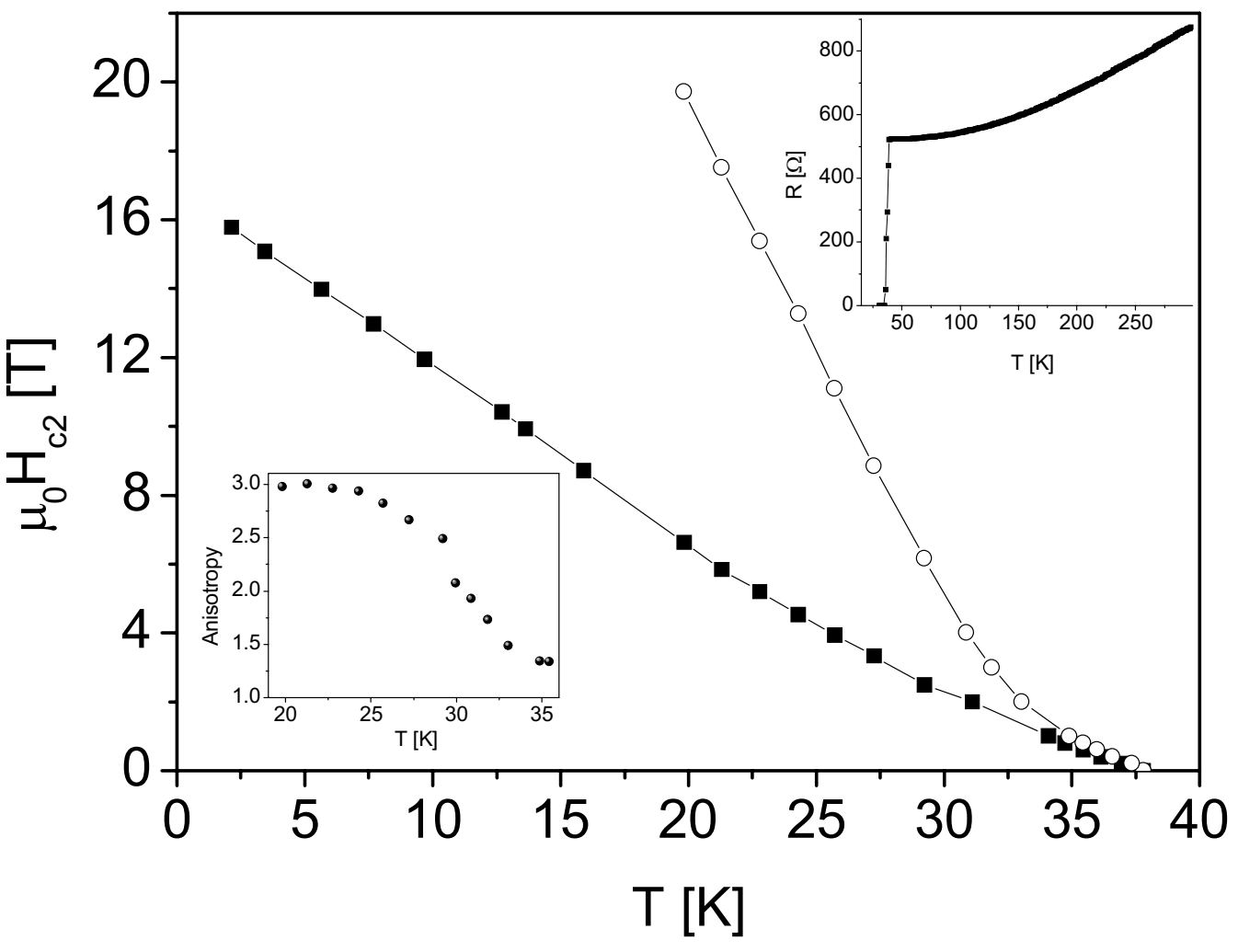

Figure 8